\documentstyle[12pt]{article}
\begin{document}
\topmargin -0.2cm \oddsidemargin -0.2cm \evensidemargin -1cm
\textheight 22cm \textwidth 12cm

\title{Creation of Electron Spinless Pairs in the Superconductivity.}  
\author{Minasyan V.N. \\
Yerevan, Armenia}

\date{\today}

\maketitle

\begin{abstract} 
First, it is demonstrated that the Froolich Hamiltonian of system in 
the superconductivity, proposed by the model of a phonon gas and an electron gas 
mixture, contains a subtle error. In this respect, we present a correct form of 
the Froolich Hamiltonian of system where the term of the interaction between the 
phonon modes and the density modes of the electron modes is described by the term o
f scattering, introduced by the Froolich in a phonon gas electron gas mixture. 
The later is removed by a canonical transformation of the Froolich Hamiltonian 
by an appearance of the attractive interaction mediated via the electron modes, w
hich leads to a bound state on a spinless electron pairs. In this letter, we show 
that the Cooper approximation as the constancy of the density states within around 
of the Fermi level has a flaw because the effective attractive forces cannot create 
the Cooper pairs into energetic gap at the Fermi level. In this letter, we find a 
condition for density metal which determines metal as superconductor.  
\end{abstract} 


\vspace{100mm}

\vspace{5mm} 
 
{\bf 1. Introduction.} 
 
\vspace{5mm} 
It is well known a lot various models [1-3] for theoretical 
investigations of the superconductivity. In 1928, Sommerfeld-Bloch had 
presented the model of individual electrons, which are moved in homogenous 
positive background of ions lattice, which considers the given system 
as electro-neutral. This model gives a good describing of the property 
metals, but it cannot explain superconductivity. Therefore, to solve the 
problem of superconductivity, the Fr$\ddot o$lich had proposed to take 
into consideration the action of  a vibration of ions lattice on a moving of free 
electrons. The later creates the effective attractive forces between 
electron modes as well as phonon modes [4]. In this letter, we show that the Fr$\ddot o$lich Hamiltonian contains a subtle error. The correction form of later leads to creation of the term effective interaction only between electron modes but not between phonon modes as it is stated by the Fr$\ddot o$lich [4]. Hence, we also remark the theory of superconductor, presented by Bardeen, Cooper and Schrieffer [5], and by Bogoliubov [6] (BCSB), which asserts that at 
the Fermi level, the Fr$\ddot o$lich effective attractive potential between 
electrons leads to shaping of two electrons with opposite spins into 
the Cooper pairs [7].  In this letter, it is demonstrated that Cooper pairs cannot be created by action an effective attractive forces between electrons around Fermi 
level. However, the model of a phonon gas -electron gas mixture for superconductor, due to an application canonical transformation of the correct form Froolich Hamiltonian of system, the term of the interaction between the density of the phonon modes and the density of the electron modes is removed by meditated an effective attractive interaction between the electron modes, which in turn determines a bound state on spinless electron pair.

\vspace{5mm}
{\bf 2. Analysis.}
\vspace{5mm}
 
We now attempt to describe the thermodynamic property of the model a phonon-electron gas mixture confined in a box of volume $V$.  In this context, we consider a electron gas as an ideal Fermi gas consisting of  $n$ free electrons with mass $m_e$ which interact with the phonon modes of the lattice by the constancy interaction which was introduced by the Froolich [4]. The main part of the Froolich Hamiltonian consists of the term of the Hamiltonian of the phonon excitations and the term of the Hamiltonian of an ideal Fermi electron gas as well as the term of the interaction between the density of the phonon excitations and the density of the electron modes: 

\begin{equation}
\hat{H}=\sum_{\vec{k},\sigma }\varepsilon_{\vec{k}}
\hat{a}^{+}_{\vec{k},\sigma }\hat{a}_{\vec{k},\sigma} +
\sum_{\vec{w}}\hbar w s\hat{b}^{+}_{\vec{w}}
\hat{b}_{\vec{w}}+i\sum_{\vec{w},\vec{k}}D_w 
\biggl(\hat{b}_{\vec{w}}\hat{\varrho}^{+}_{\vec{w}}-
\hat{b}^{+}_{\vec{w}}\hat{\varrho}_{\vec{w}}\biggl)
\end{equation}

where  $\hat{a}^{+}_{\vec{k},\sigma}$ and 
$\hat{a}_{\vec{k},\sigma }$ are, respectively,  the Fermi operators of creation and 
annihilation for free electron with wave-vector $\vec{k}$ and energy $ \varepsilon_{\vec{k}}=\frac{\hbar^2 k^2}{2m_e}$, by the value of its 
spin z-component $\sigma=^{+}_{-}\frac{1}{2}$; $s$ is the velocity of phonon; $\hat{b}^{+}_{\vec{w},\sigma}$ and $\hat{b}_{\vec{w},\sigma }$ are, respectively,  the Bose operators of creation and annihilation for free phonon with wave-vector $\vec{w}$ and energy$\hbar w s$;  $D_w$ is the constant of the interaction between  the density of the phonon excitations and the density modes of the electrons which equals to $ D_w =\sqrt{\frac{\alpha \hbar w s}{V} }$ (where $\alpha=\frac{{C^{"}}^2}{2M s^2\frac{n}{V}}$ is the constant charactering of a metal; $ C^{"}$ is the constant of the interaction; $M$ is the mass of ion) ; $\hat{\varrho}_{\vec{w}}$ is the density operator of the electron modes with wave vector  $\vec{w}$ which 
is defined as:
\begin{equation}
\hat{\varrho}_{\vec{w}}=\sum_{\vec{k}, \sigma }
\hat{a}^{+}_{\vec{k}-\vec{w},\sigma }\hat{a}_{\vec{k},\sigma }
\end{equation}
and 

\begin{equation}
\hat{\varrho}^{+}_{\vec{w}}=\sum_{\vec{k}, \sigma }
\hat{a}^{+}_{\vec{k},\sigma }\hat{a}_{\vec{k}-\vec{w},\sigma }
\end{equation}

where $\hat{\varrho}^{+}_{\vec{w}}=\hat{\varrho}_{-\vec{w}}$

Hence, we note that the Fermi operators $\hat{a}^{+}_{\vec{k},\sigma}$ and $\hat{a}_{\vec{k},\sigma }$ satisfy to the Fermi commutation relations $[\cdot\cdot\cdot]_{+}$ as:

\begin{equation}
\biggl[\hat{a}_{\vec{k},\sigma}, \hat{k}^{+}_{\vec{p}^{'},
\sigma^{'}}\biggl]_{+} =
\delta_{\vec{k},\vec{k^{'}}}\cdot\delta_{\sigma,\sigma^{'}}
\end{equation}

\begin{equation}
[\hat{a}_{\vec{k},\sigma}, \hat{a}_{\vec{k^{'}}, \sigma^{'}}]_{+}= 0
\end{equation}

\begin{equation}
[\hat{a}^{+}_{\vec{k},\sigma}, \hat{k}^{+}_{\vec{p^{'}}, 
\sigma^{'}}]_{+}= 0
\end{equation}

On other hand, the Bose operators $\hat{b}^{+}_{\vec{w}}$ and 
$\hat{b}_{\vec{w}}$ satisfy to the Bose commutation relations 
$(\cdot\cdot\cdot)$ as:

\begin{equation}
\biggl(\hat{b}_{\vec{w}}, \hat{b}^{+}_{\vec{w^{'}}}\bigg) =
\delta_{\vec{w},\vec{w^{'}}}
\end{equation}

\begin{equation}
\biggl(\hat{b}_{\vec{w}}, \hat{b}_{\vec{w^{'}}}\biggl)= 0
\end{equation}

\begin{equation}
\biggl(\hat{b}^{+}_{\vec{w}}, \hat{b}^{+}_{\vec{w^{'}}}\biggl)= 0
\end{equation}

Obviously, the Bose- operator  $\hat{b}_{\vec{w}}$ 
commutates with  the Fermi operator $\hat{a}_{\vec{k},\sigma}$, which it means 
that the phonon excitations and electron modes are an independent.

We now show that the Hamiltonian of system $\hat{H}$ in (1), proposed by the Fr$\ddot o$lich at solving of the problem superconductivity (please, see the Equation (15) in H. Fr$\ddot o$lich, Proc.Roy. Soc, {\bf A215}, 291-291 (1952) in the reference [4] ), contains a subtle error in the term of the interaction between the density of phonon modes and the density of electron modes which represents a third term in right side of (1). Indeed, the later is described by two sums, one from which goes by a wave vector $\vec{w}$ but other sum goes through by a wave vector  $\vec{k}$. This fact contradicts to the definition of the density operator of the electron modes  $\hat{\varrho}_{\vec{w}}$ (2) which in turn already contains the sum by the wave vector $\vec{k}$, and therefore, it is not a necessary to take into account so-called twice summation from $\vec{k}$ and $\vec{w}$ for describing of the term of interaction between the density of phonon modes and the density of electron modes. Thus, the correct form of the Fr$\ddot o$lich Hamiltonian, due to the definition of the density operator of the electron modes $\hat{\varrho}_{\vec{w}}$, must be presented as: 

\begin{equation}
\hat{H}=\hat{H}_0 +\hat{H}_1 +\hat{H}_2
\end{equation}

where

\begin{equation}
\hat{H}_0=\sum_{\vec{k},\sigma }\varepsilon_{\vec{k}}
\hat{a}^{+}_{\vec{k},\sigma }\hat{a}_{\vec{k},\sigma} 
\end{equation}

\begin{equation}
\hat{H}_1=\sum_{\vec{w}}\hbar w s\hat{b}^{+}_{\vec{w}}
\hat{b}_{\vec{w}}
\end{equation}

\begin{equation}
\hat{H}_2=i\sum_{\vec{w}}D_w \biggl(\hat{b}_{\vec{w}}\hat{\varrho}^{+}_{\vec{w}}-\hat{b}^{+}_{\vec{w}}\hat{\varrho}_{\vec{w}}\biggl) 
\end{equation}
The later defines the correct form of the Fr$\ddot o$lich Hamiltonian. However, hence, we introduce new transformation of the Bose -operators of phonon modes $\hat{b}^{+}_{\vec{w}}$ and $\hat{b}_{\vec{w}}$ by the new Bose –operators of phonon excitations $\hat{c}^{+}_{\vec{w}}$ and $\hat{c}_{\vec{w}}$ which help us to remove an anomalous term (13) in (10): 

\begin{equation}
\hat{b}_{\vec{w}}=-i\hat{c}_{\vec{w}}
\end{equation}
and

\begin{equation}
\hat{b}^{+}_{\vec{w}}=i\hat{c}^{+}_{\vec{w}}
\end{equation}
 
Then (12) and  (13) in the (10) take a following forms:

\begin{equation}
\hat{H}_1=\sum_{\vec{w}}\hbar w s\hat{c}^{+}_{\vec{w}}
\hat{c}_{\vec{w}}
\end{equation}

\begin{equation}
\hat{H}_2=\sum_{\vec{w}}D_w \biggl(\hat{c}_{\vec{w}}\hat{\varrho}^{+}_{\vec{w}}+\hat{c}^{+}_{\vec{w}}\hat{\varrho}_{\vec{w}}\biggl) =\sum_{\vec{w}}D_w\hat{\varrho}_{\vec{w}}\biggl(\hat{c}_{\vec{w}}+
\hat{c}^{+}_{\vec{w}}\biggl)
\end{equation}

To allocate anomalous term in the Hamiltonian of system $\hat{H}$ in (10),  presented by the term in (13), the Fr$\ddot o$lich claims to use of a following  approach which allows to do a canonical transformation for
the operator $\hat{H}$. Letting that the operator $\hat{S}$ satisfies to the condition $\hat{S}^{+} = -\hat{S}$ within introducing of new operator $\tilde{H}$:

\begin{equation}
\tilde{H}=\exp\biggl(\hat{S}^{+}\biggl)\hat{H}
\exp\biggl(\hat{S}\biggl)
\end{equation}

which is decayed  by the following terms: 

\begin{equation}
\tilde{H}=\exp\biggl(\hat{S}^{+}\biggl) \hat{H}
\exp\biggl(\hat{S}\biggl) = \hat{H}-(\hat{S},\hat{H})+
\frac{1}{2}(\hat{S},(\hat{S},\hat{H}))-\cdots
\end{equation}

At least, we consider the Frolich introductions of following determinations as 
   
\begin{equation}
\hat{S}=\sum_{\vec{w}}\hat{S_{\vec{w}}}
\end{equation}
with the operator $\hat{S_{\vec{w}}}$ presented by terms of new Bose-operators

\begin{equation}
\hat{S_{\vec{w}}}=-\gamma_{\vec{w}}
\hat{b}_{\vec{w}}+
\gamma^{+}_{\vec{w}}\hat{b}^{+}_{\vec{w}}=i\gamma_{\vec{w}}
\hat{c}_{\vec{w}}+
i\gamma^{+}_{\vec{w}}\hat{c}^{+}_{\vec{w}}
\end{equation}
where 

\begin{equation}
\gamma_{\vec{w}}=\sum_{\vec{k}, \sigma }
\varphi (\vec{k},\vec{w})\hat{a}^{+}_{\vec{k},\sigma }\hat{a}_{\vec{k}-\vec{w},\sigma }
\end{equation}
and

\begin{equation}
\gamma^{+}_{\vec{w}}=\sum_{\vec{k}, \sigma }
\varphi^{+} (\vec{k},\vec{w})
\hat{a}^{+}_{\vec{k}-\vec{w},\sigma }
\hat{a}_{\vec{k},\sigma }
\end{equation}
where $\varphi (\vec{k},\vec{w})$ is the c-number [4]. 

The fact, as removing of a subtle error in the Frolich Hamiltonian, claims
to  suggest that  $\varphi (\vec{k},\vec{w})=\varphi^{+} (\vec{k},\vec{w})=\varphi (\vec{w})$ where $\varphi (\vec{w})=\varphi (-\vec{w})$. Then, the operators $\gamma_{\vec{w}}$ and $\gamma^{+}_{\vec{w}}$ take  the following forms:

\begin{equation}
\gamma_{\vec{w}}= \varphi (\vec{w})\sum_{\vec{k}, \sigma }
\hat{a}^{+}_{\vec{k},\sigma }\hat{a}_{\vec{k}-\vec{w},\sigma }= \varphi (\vec{w})\hat{\varrho}^{+}_{\vec{w}}
\end{equation}
and

\begin{equation}
\gamma^{+}_{\vec{w}}=\varphi (\vec{w})\sum_{\vec{k}, \sigma }
\hat{a}^{+}_{\vec{k}-\vec{w},\sigma }
\hat{a}_{\vec{k},\sigma }=\varphi (\vec{w})\hat{\varrho}_{\vec{w}}
\end{equation}
which by inserting into (21) with using of (20), we obtain
\begin{equation}
\hat{S}=\sum_{\vec{w}}\hat{S_{\vec{w}}}= \sum_{\vec{w}}\varphi (\vec{w})\hat{\varrho}_{\vec{w}}\biggl (
\hat{c}_{-\vec{w}}-\hat{c}^{+}_{\vec{w}}\biggl)
\end{equation}
at application of conditions $\hat{\varrho}^{+}_{-\vec{w}}=\hat{\varrho}_{\vec{w}}$ and $\varphi (-\vec{w})=\varphi (\vec{w})$ 

In analogy manner, 

\begin{equation}
\hat{S}^{+}=\sum_{\vec{w}}\hat{S^{+}_{\vec{w}}}=
\sum_{\vec{w}}\varphi (\vec{w})\hat{\varrho}^{+}_{\vec{w}}\biggl (
\hat{c}^{+}_{-\vec{w}}-\hat{c}_{\vec{w}}\biggl)=
-\sum_{\vec{w}}\varphi (\vec{w})\hat{\varrho}_{\vec{w}}\biggl (
\hat{c}_{-\vec{w}}-\hat{c}^{+}_{\vec{w}}\biggl)
\end{equation}

For further our calculations, it needs to extend the form of Equation (16) and (17) in [4] within application $\hat{H}_0$ from (11) and $\hat{\varrho}_{\vec{w}}$ from (2), and then, we have

\begin{equation}
\biggl(\gamma_{\vec{w}}, \hat{H}_0\biggl)=- \varphi (\vec{w} ) \biggl(\sum_{\vec{k},\sigma }\varepsilon_{\vec{k}}\hat{a}^{+}_{\vec{k},\sigma }\hat{a}_{\vec{k}-\vec{w},\sigma }-\sum_{\vec{k},\sigma }\varepsilon_{\vec{k}+\vec{w}}\hat{a}^{+}_{\vec{k}+\vec{w},\sigma }\hat{a}_{\vec{k},\sigma }\biggl)=0
\end{equation}
and
\begin{equation}
\biggl(\gamma_{\vec{w}},\hat{\varrho}_{\vec{w}}\biggl)=
\biggl(\gamma_{\vec{w}}, \sum_{\vec{k},\sigma }
\hat{a}^{+}_{\vec{k}-\vec{w},\sigma }
\hat{a}_{\vec{k},\sigma }\biggl)=- \varphi (\vec{w} ) \biggl (\sum_{\vec{k},\sigma } \hat{n}_{\vec{k}-\vec{w},\sigma } -\sum_{\vec{k},\sigma } \hat{n}_{\vec{k},\sigma }\biggl)=0
\end{equation}

because $\sum_{\vec{k},\sigma }\hat{n}_{\vec{k}-\vec{w},\sigma }=\sum_{\vec{k},\sigma } \hat{n}_{\vec{k},\sigma }$.

On other hand, using of the commutation expressions:

\begin{equation}
\biggl(\hat{c}_{\vec{w}}, \hat{c}^{+}_{\vec{w}^{'}}
\hat{c}_{\vec{w}^{'}}\bigg)]_{+} =
\hat{c}_{\vec{w}}\delta_{\vec{w},\vec{w}^{'}}
\end{equation}
and

\begin{equation}
\biggl(\hat{c}^{+}_{\vec{w}}, \hat{c}^{+}_{\vec{w}^{'}}
\hat{c}_{\vec{w}^{'}}\bigg) =
-\hat{c}^{+}_{\vec{w}}\delta_{\vec{w},\vec{w}^{'}}
\end{equation}
We may calculate the commutation terms included in (19) by substitutions (26), (28) and (29) into (19). Thus, we arrive 

\begin{equation}
\biggl (\hat{S},\hat{H}\biggl)= \biggl (\hat{S},\hat{H}_0\biggl)+ \biggl (\hat{S},\hat{H}_1\biggl)+ \biggl (\hat{S},\hat{H}_2\biggl)
\end{equation}

where

\begin{equation}
\biggl (\hat{S},\hat{H}_0\biggl)= 0
\end{equation}

\begin{equation}
\biggl (\hat{S},\hat{H}_1\biggl)= 
\sum_{\vec{w}} \varphi (\vec{w}) \hbar w s\hat{\varrho}_{\vec{w}} 
\biggl(\hat{c}_{-\vec{w}}+\hat{c}^{+}_{\vec{w}}\biggl)
\end{equation}

\begin{equation}
\biggl (\hat{S},\hat{H}_2\biggl)= 2\sum_{\vec{w}} 
\varphi (\vec{w}) D_w\hat{\varrho}_{\vec{w}}
\hat{\varrho}_{-\vec{w}} 
\end{equation}
Thus,

\begin{equation}
\biggl (\hat{S},\hat{H}\biggl)= sum_{\vec{w}} \varphi (\vec{w}) \hbar w s\hat{\varrho}_{\vec{w}} 
\biggl(\hat{c}_{-\vec{w}}+\hat{c}^{+}_{\vec{w}}\biggl)+
2 \sum_{\vec{w}} 
\varphi (\vec{w}) D_w\hat{\varrho}_{\vec{w}}
\hat{\varrho}_{-\vec{w}} 
\end{equation}

In this respect, it is easy to show that
\begin{equation}
\frac{1}{2}\biggl (\hat{S},\biggl (\hat{S},\hat{H}\biggl) \biggl )=-\sum_{\vec{w}}\varphi^2 (\vec{w})  \hbar w s\varrho_{\vec{w}}
\hat{\varrho}_{-\vec{w}}
\end{equation}

and $\biggl(\hat{S}, \biggl(\hat{S},\biggl(\hat{S},\hat{H}\biggl)\biggl)\biggl)=0$

In this context, the new operator $\tilde{H}$ in (19) takes a following form:

\begin{eqnarray}
\tilde{H}& =&\sum_{\vec{k},\sigma }\varepsilon_{\vec{k}}
\hat{a}^{+}_{\vec{k},\sigma }\hat{a}_{\vec{k},\sigma} +
\sum_{\vec{w}}\hbar w s\hat{c}^{+}_{\vec{w}}
\hat{c}_{\vec{w}}+\sum_{\vec{w}}D_w 
\biggl(\hat{c}_{\vec{w}}+
\hat{c}^{+}_{-\vec{w}}\biggl) \hat{\varrho}_{-\vec{w}}-
\nonumber\\
&-&\sum_{\vec{w}} \varphi (\vec{w}) \hbar w s \hat{\varrho}_{\vec{w}}
\varphi (\vec{w})\biggl(\hat{c}_{-\vec{w}}+\hat{c}^{+}_{\vec{w}}\biggl)
-2 \sum_{\vec{w}} 
\varphi (\vec{w}) D_w\hat{\varrho}_{\vec{w}}
\hat{\varrho}_{-\vec{w}} +
\nonumber\\
&+&\sum_{\vec{w}}\varphi^2 (\vec{w})  \hbar w s\varrho_{\vec{w}}\hat{\varrho}_{-\vec{w}}
\end{eqnarray}

To find a value of $\varphi (\vec{w})$, we takes that the sum of a  third and a fourth terms, in right side of (38), is zero:

\begin{equation}
\varphi (\vec{w})=\frac{D_w }{\hbar w s }
\end{equation}

In this respect, the new Hamiltonian of system (38) arrives to following form:

\begin{eqnarray}
\tilde{H}& =&\sum_{\vec{w}}\hbar w s\hat{c}^{+}_{\vec{w}}
\hat{c}_{\vec{w}}+\sum_{\vec{k},\sigma }\varepsilon_{\vec{k}}
\hat{a}^{+}_{\vec{k},\sigma }\hat{a}_{\vec{k},\sigma}-
2 \sum_{\vec{w}} 
\varphi (\vec{w}) D_w\hat{\varrho}_{\vec{w}}
\hat{\varrho}_{-\vec{w}}  +
\nonumber\\
&+&\sum_{\vec{w}}\varphi^2 (\vec{w})  \hbar w s\varrho_{\vec{w}}\hat{\varrho}_{-\vec{w}}
\end{eqnarray}
which by inserting $\varphi (\vec{w})$ from (39) into (38), and with taking into account (14) and (15), may present by a form:

\begin{equation}
\tilde{H}= \sum_{\vec{w}}\hbar w s\hat{b}^{+}_{\vec{w}}
\hat{b}_{\vec{w}}+\sum_{\vec{k},\sigma }\varepsilon_{\vec{k}}
\hat{a}^{+}_{\vec{k},\sigma }\hat{a}_{\vec{k},\sigma}
- \sum_{\vec{w}} \frac{ D^2_w}{\hbar w s }\hat{\varrho}_{\vec{w}}
\hat{\varrho}_{-\vec{w}} 
\end{equation}

where denotes of  the effective Hamiltonian of  an electron gas $\hat{H}_e $ which contains an effective interaction between neutron modes, and then we may rewrite down:

\begin{equation}
\tilde{H}= \sum_{\vec{w}}\hbar w s\hat{b}^{+}_{\vec{w}}
\hat{b}_{\vec{w}}+\hat{H}_e
\end{equation}
with 

\begin{equation}
\hat{H}_e =\sum_{\vec{k},\sigma }\varepsilon_{\vec{k}}
\hat{a}^{+}_{\vec{k},\sigma }\hat{a}_{\vec{k},\sigma} +\frac{1}{2V}\sum_{\vec{w}}V_{\vec{w}}
\hat{\varrho}_{\vec{w}}\hat{\varrho}_{-\vec{w}}
\end{equation}

where $V_{\vec{w}}$ is the effective potential of the interaction between electron modes, which at taking into account $ D_w =\sqrt{\frac{\alpha \hbar w s}{V} }$, has the form:

\begin{equation}
V_{\vec{w}}= -\frac{2D^2_w V}{\hbar w s } =-2\alpha  
\end{equation}

Consequently, the effective scattering between two electrons is presented in the coordinate space by a following form:

\begin{equation}
V (\vec{r})=\frac{1}{V}\sum_{\vec{w}} V_{\vec{w}}\cdot e^{i\vec{w}\vec{r}}=-2\alpha\delta (\vec{r})
\end{equation}

because it is well known that $\frac{1}{V}\sum_{\vec{w}} e^{i\vec{w}\vec{r}}=\delta (\vec{r})$.

\vspace{5mm}
{\bf 3. The subtle error in the Cooper theory.}
\vspace{5mm}

In 1956 Cooper had suggested that the density of states around the 
Fermi level could be approximated by its value at the Fermi energy 
$E_f$  [7]. To find own meaning of the binding energy $E_0<0$ of the electron pair system in the ground state, we use of the Schr$\ddot o$dinger equation from the Cooper theory with denotes presented in this letter:

\begin{equation}
1=\frac{2\alpha m_e }{V}\sum_{E_f <E_p<E_f+\delta} \frac{1}{E_p-E_0}
\end{equation}
where $E_p=\frac{p^2}{m_e}$; $E_f=\frac{p^2_f}{2m_e}$ is the Fermi energy; $p_f$ is the momentum of Fermi; $\delta $ is the thickness of the energetic layer which is a very small in respect to Fermi energy $\delta \ll E_f$, at the Fermi level.

The solution of (46) for finding of the binding energy $E_0<0$ suggests to calculate of right part of (46):

\begin{equation}
\frac{1}{V}\sum _{E_f <E_p<E_f+\delta} \frac{1}{ E_p -E_0}=\frac{1}{2\pi\hbar^3}\int^{p_f+\frac{\delta m_e}{p_f}}_{p_f} \frac{ p^2}{\frac{p^2}{m_e}-E_0}d p =A
\end{equation}
At application of the well known mathematical formulae 
$$
\int \frac{x^2dx}{x^2-a^2}=x-\frac{a}{2}ln|\frac{a+x}{a-x}|
$$

and using of it into (46), we obtain 

\begin{equation}
1=\frac{\alpha m_e A}{\pi^2\hbar^3} 
\end{equation}

where

\begin{equation}
A=\frac{\delta m_e}{p_f}-\frac{\sqrt{-E_0 m_e}}{2}\cdot 
ln\mid\frac{p_f+\frac{\delta m_e}{p_f}+\sqrt{-m_e E}}
{p_f+\frac{\delta m_e}{p_f}-\sqrt{-m_e E_0}}\mid 
\end{equation}

Obviously, the value of second term in right side of (49) is  
$$
\frac{\sqrt{-E_0 m_e}}{2}\cdot 
ln\mid\frac{p_f+\frac{\delta m_e}{p_f}+\sqrt{-m_e E}}
{p_f+\frac{\delta m_e}{p_f}-\sqrt{-m_e E_0}}\mid >0
$$  
therefore, as result of (49), $ A<\frac{\delta m_e}{p_f}$, which replaces the Eq.(48) as an inequality:   

\begin{equation}
1=\frac{\alpha m_e A}{\pi^2\hbar^3}<\frac{\alpha \delta m^2_e }{\pi^2\hbar^3 p_f} 
\end{equation}

Hence, we note that the scattering term $V_w$ in (44) can be expressed by term of S-wave potential interaction between two electrons [6]: 
\begin{equation}
V_w=-\frac{4\pi\hbar^2 d}{m_e}=-2\alpha
\end{equation}
where $d$ is the scattering amplitude which in turn satisfies to a 
condition $\frac{p_f d}{\hbar}\ll 1$. The later condition is a very important factor for existence of the term of interaction between phonon excitations and electron modes by S-wave scattering, therefore, inserting value $\alpha $ from (51) to (50), we have

\begin{equation}
1<\frac{2 \delta m_e}{\pi p^2_f}\cdot\frac{d p_f}{\hbar}\ll 
\frac{1}{\pi}\frac{\delta }{E_f}
\end{equation}
at $\frac{p_f d}{\hbar}\ll 1$ and $E_f=\frac{ p^2_f }{2m_e}$.

It is a well known for example that the ratio $\frac{\delta }{E_f}\approx 0.1$. This fact implies that the right side of (52) equals to $\frac{1}{\pi}\frac{\delta }{E_f}=\frac{0.1}{\pi}=0.03$, and then, the inequality (52) represents as a non-sense result because there is a non sense inequality $1\ll 0.03$. Consequently, we can state that the attractive potential cannot couple two electrons as the Cooper pair because (48) cannot have a solution at condition $\frac{p_f d}{\hbar}\ll 1$. 

We now show an origin of creation of so-called flaw in the theory of Cooper. As it is known, for finding of  meaning  $A$ in Eq.(47), the Cooper introduced an approximation for density electron states $\varrho (E)$ which suggest a following approximatiom as $\varrho (E)\approx \varrho(E_f)$. In this respect, the Cooper calculated a value of $A$ in  (49) by following way:

\begin{equation}
A=\frac{1}{V}\sum _{E_f <E<E_f+\delta} \frac{1}{E-E_0}=\int^{E_f +\delta}_{E_f}\frac{\varrho (E) dE}{E-E_0}=\varrho (E_f) ln\mid \frac{E_f +\delta -E_0}{ E_f -E_0}\mid>0
\end{equation}
where a value of $A>0$ is a positive number. In this context, the equation (48) may have a solution, and always is fulfilled by support of a wrong approximation $\varrho (E)\approx \varrho(E_f)$ which was introduced by Cooper by handling. However, as we have been seen, the correct form of value $A$, presented in (48), leads to a non-sense result.  

To solve the problem connected with creation electron pair we try to replace the strong interaction between two electrons in (45) by a screening effective scattering.

\vspace{5mm}
{\bf 4. Creation Spinless Electron Pairs.}
\vspace{5mm}

The term of the interaction between two electrons $V (\vec{r})$ in the coordinate space mediates the attractive Coulomb interaction between two charged particles with mass of electron $m_e$, having the opposite effective charges $ e_*$ and $- e_*$, which together create a neutral system. Indeed, the effective Hamiltonian of an electron gas is rewritten down in the space of coordinate by following form:

\begin{equation}
\hat{H}_n =\sum^{\frac{n}{2}}_{i=1}\hat{H}_i =-\frac{\hbar^2}{2m_e}\sum^{n}_{i=1} \Delta_I +\sum_{i<j} V (\mid\vec{r}_i-\vec{r}_j\mid)
\end{equation}

where $\hat{H}_i $ is the Hamiltonian of system consisting two electron with opposite spin which have a coordinates $\vec{r}_i $ and $\vec{r}_j $:
\begin{equation}
\hat{H}_i=-\frac{\hbar^2}{2m_e}\Delta_i-\frac{\hbar^2}{2m_e}\Delta_j +V (\mid\vec{r}_i-\vec{r}_j\mid)
\end{equation}
The transformation of considering coordinate system to the relative coordinate $\vec{r}=\vec{r}_i-\vec{r}_j $ and the coordinate of center mass $\vec{R}=\frac{\vec{r}_i+\vec{r}_j }{2}$, claims 

\begin{equation}
\hat{H}_i=-\frac{\hbar^2}{4m_e}\Delta_R-\frac{\hbar^2}{m_e}\Delta_r +V (\vec{r})
\end{equation}
To find the binding energy $E<0$ of electron pair, we search the solution of a wave equation with introduction of wave function $\psi(\vec{r})$ by presentation of the Schr$\ddot o$dinger equation:
$$
\hat{H}_i \psi_s (\vec{r})=E\psi_s (\vec{r})
$$
In this respect, we have a following equation 

\begin{equation}
-\frac{\hbar^2}{m_e}\Delta_r\psi_s (\vec{r})+ V (\vec{r})\psi_s (\vec{r}) = E\psi(\vec{r})
\end{equation}
which may determine the binding energy $E<0$ of electron pair, if we claim that the  
condition $\frac{p_f d}{\hbar}\ll 1$ always is fulfilled. This reasoning implies that 
the effective scattering between two electrons in (45) is presented in the coordinate space by a well known following form:

\begin{equation}
V (\vec{r})=\frac{1}{V}\sum_{\vec{w}} V_{\vec{w}}\cdot e^{i\vec{w}\vec{r}}=4\pi\int^{w_f}_{0} V_{\vec{w}} w^2\frac{sin (w r)}{w r}d w
\end{equation}
where we introduce a following approximation as $ \frac{sin (w r)}{w r}\approx 1-\frac{w^2r^2}{6}$ because the conditions $w\leq w_f$ and $w_f d\ll 1$ lead to $ w_f r \ll 1$  ($w_f=\biggl(\frac{3\pi^2 n}{V}\biggl)^{\frac{1}{3}}$ is the Fermi wave number). The later condition defines a state for distance $r$ between two neighboring electrons $r\ll \frac{1}{w_f}=\biggl(\frac{V}{3\pi^2 n}\biggl)^{\frac{1}{3}}$. In this context, by taking into consideration $\frac{4\pi w^3_f}{3}=\frac{n}{2V}$ and (44), we obtain    

\begin{equation}
V (\vec{r})\approx- \frac{\alpha n}{V} +\alpha \biggl(\frac{n}{V}\biggl)^{\frac{5}{3}}\cdot r^2
\end{equation}

Thus, the effective interaction between electron modes $ V (\vec{r})=-2\alpha \delta (\vec{r})$ (45) is replaced by a screening effective scattering presented in (59). This approximation means that there is an appearance of a screening character in the effective scattering because as we see the later depends on the density electron modes. Inserting value of $ V (\vec{r})$ from (59) to (57), and denoting $E=E_s$

\begin{equation}
\biggl[-\frac{\hbar^2}{m_e}\Delta_{r}- \frac{n\alpha }{V}+\alpha \biggl(\frac{n}{V}\biggl)^{\frac{5}{3}}\cdot r^2\biggl]
\psi_s (r)=E_s \psi_s (r)
\end{equation}

We transform the form (60) by a following form:

\begin{equation}
\frac {d^2 \psi_s (r)}{d r^2} + \biggl (\lambda-\theta ^2 r^2 \biggl ) \psi_s (r)=0
\end{equation}

where we take $\theta=-\sqrt{\frac{m_e\alpha}{\hbar^2} \biggl(\frac{n}{V}\biggl)^{\frac{5}{3}}}$, and $\lambda =\frac{ m_e E_s}{\hbar^2}-\frac{\alpha m_e  n }{\hbar^2V}$

By application of the wave function 
$\psi_s (r)$ via the Chebishev-Hermit function 
$H_s (it)$ from an imaginary number as argument  $it$ [8] (where 
$i$ is the imaginary one; $t$ is the real number; $s=0;1;2;\cdots$), the equation (61) has a following solution as:

$$
\psi_s(\vec{r})=e^{-\theta\cdot r^2}H_s(\sqrt{\theta }\cdot r) 
$$
where 
$$
H_s(it)=i^s e^{-t^2}\frac{d^s e^{t^2}}{d t^s}
$$ 
at $\theta<0$ within
$$
\lambda=\theta (s+\frac{1}{2})
$$

Consequently, the quantity of the binding energy $E_s$ of electron pair with mass $m_0=2m_e$ is rewritten as: 
\begin{equation}
E_s =-\sqrt{ \frac{\alpha\hbar^2}{m_e} \biggl(\frac{n}{V}\biggl)^{\frac{5}{3}}}\biggl(s+\frac{1}{2}\biggl) +\frac{\alpha  n }{V}< 0
\end{equation}
at $s=0;1;2;\cdots$

The normal state of electron pair corresponds to quantity  $s=0$ which defines the maximal quantity of the binding energy of ion pair: 
\begin{equation}
E_0 =-\sqrt{\frac{\alpha\hbar^2}{m_e}\biggl(\frac{n}{V}\biggl)^{\frac{5}{3}}}+\frac{\alpha n }{V} <0
\end{equation}
which implies that the creation of the superconducting phase of the metal is appeared by the condition for density of metal $\frac{n}{V}$:
$$
\frac{n}{V}>\biggl(\frac{C^2 m_e }{2M s^2\hbar^2}\biggl)^{\frac{3}{2}}
$$
At choosing $C\approx 10ev$ [4]; $M\approx 5\cdot 10^{-26} kg$; $s\approx 3\cdot 10^3 m$, we estimate that any metal with density of electron 
$$
\frac{n}{V}>10^{27}\frac{1}{m^3}
$$ 
may  represent as a superconductor.

Thus, the spinless electron pair is created in a phonon gas-electron gas mixture by the term of the interaction between the phonon excitations and electron modes which in turn is removed by an induced the effective interaction which mediate via electron modes. The later determines a bound state on 
a electron pair with binding energy (63) which depends on the density of electron modes $\frac{n}{V}$, the constant of interaction $C$ and mass of ion $M$. In accordance with this reasoning, the Hamiltonian system with superconducting phase takes a following form:

\begin{equation}
\tilde{H}= \sum_{\vec{p}}\varepsilon_{\vec{p}}
\hat{b}^{+}_{\vec{p}}\hat{b}_{\vec{p}}+\sum_{\vec{p}}\frac{p^2}{2m_0}
\hat{d}^{+}_{\vec{p}}\hat{d}_{\vec{p}} 
\end{equation}
where $\hat{d}^{+}_{\vec{p}}$ and $\hat{d}_{\vec{p}}$ are, respectively, the
"creation" and  "annihilation" Bose-operators of a free electron pairs with 
momentum $\vec{p}$.

In conclusion, we note that the new solution of the problem superconductivity, 
presented in this letter, might be useful for describing 
of the thermodynamic properties of a dilute gas of the Boson-Fermion mixtures confined in traps. This fact implies that the Fr$\ddot o$lich could discover the electron pairs in a superconductivity earlier then it was made by the Cooper [7] by support of wrong approximation.

\newpage 
\begin{center} 
{\bf References} 
\end{center}

\begin{enumerate}
\item
P.W.~Anderson~., The theory of superconductivity in the high
-$T_c$ cuprates (Prinston University Press,Princeton, New Jersey,
1997).
\item
A.S.~Alexandrov~and Sir Nevil ~Mott~,"Polarons  and Bipolarons",
Cambridge- Loughborough, Word Scientific, (1995).
\item
Gerald D.~Mahan~, "Many-Particle Physics", (~Plenium~ Press New
York, 1990)
\item
H.~Fr$\ddot o$lich~, Proc.Roy. Soc, {\bf A215}, 291-291 (1952)
\item
J.~Bardeen~, L.N.~Cooper~, and J.R.~Schrieffer~, Phys.Rev.{\bf
108},~1175~(1957).
\item
N.N.~Bogoliubov~, ~Nuovo~ ~Cimento~, ~{\bf 7},~794~(1958).
\item
L.N.~Cooper~, Phys.Rev., {\bf 104}, 1189 (1956).
\item 
M.A.~Lavrentiev~, and B.V. ~Shabat ~ "Nauka", ~ Moscow,~, (1973)

\end{enumerate} 
\end{document}